\documentclass[journal,draftcls,onecolumn,12pt,twoside]{IEEEtranTMBMC}
\usepackage{amsfonts, amsthm}
\usepackage{amsmath, color,hyperref}
\usepackage{blkarray}
\usepackage{graphicx}
\usepackage{cite}
\usepackage[normalem]{ulem}
\usepackage{tikz}

\usepackage[font=small, labelfont=bf]{caption}
\usetikzlibrary{arrows,automata}

\newtheorem{theorem}{Theorem}
\newtheorem{definition}{Definition}

\newtheorem{remark}{Remark}
\newtheorem{example}{Example}

\allowdisplaybreaks
\title{On ISI-free Modulations for Diffusion based Molecular Communication}
\author{Hamidreza~Arjmandi$^*$, Mohammad~Movahednasab$^*$, Amin~Gohari$^*$, Mahtab~Mirmohseni$^*$, Masoumeh~Nasiri\,Kenari$^*$,~\IEEEmembership{Member,~IEEE}, Faramarz~Fekri$^{\dagger}$,~\IEEEmembership{Senior\,Member, IEEE}\\
$^*$ Sharif University of Technology, $^{\dagger}$ Georgia Institute of Technology}
\begin{document}
\maketitle
\begin{abstract}
A diffusion molecular channel is a channel with memory, as molecules released into the medium hit the receptors after a random delay. Coding over the diffusion channel is performed by choosing the type, intensity, or the released time of molecules diffused in the environment over time. To avoid intersymbol interference (ISI), molecules of the same type should be released at time instances that are sufficiently far apart. This ensures that molecules of a previous transmission are faded in the environment, before molecules of the same type are reused for signaling.  In this paper, we consider ISI-free time-slotted modulation schemes. The maximum reliable transmission rate for these modulations is given by the constrained coding capacity of the graph that represents the permissible transmission sequences. However, achieving the constrained coding capacity requires long blocklengths and delays at the decoder, making it impractical for simple nanomachines. The main contribution of this paper is to consider modulations with small delay (short blocklength) and show that they get very close to constrained coding capacity. 
\end{abstract}
\section{Introduction}

Molecular communications (MC) via diffusion is one of the most promising approaches for communications among nanonetworks \cite{Akyl2011, Nakano2012}. In this approach, information is conveyed to a nano-receiver by nano-transmitter's choice of the concentration, type, or the release time of the molecules diffused into the medium. The motion of the released molecules are described by a Brownian motion process, that can be with or without drift. A molecule released from a transmitter can follow different trajectories before reaching a receiver. Thus, diffusion-based communication suffers from intersymbol interference (ISI) due to molecules from previous transmission that follow longer trajectories before hitting the receiver. 

The effect of ISI in diffusion based MC has been  studied extensively in the literature \cite{B2, Atakan2012, Pierobon2014, Aminian2014}, with the general conclusion that ISI reduces the performance in a communication setup consisting of one transmitter and one receiver. The authors in \cite{Kuran2011,fekri}  have proposed two modulation schemes, {\em Concentration Shift Keying (CSK)} and {\em Molecular Shift Keying (MOSK)}, where both suffer from the ISI caused by molecules from previous transmissions. Authors in \cite{Arjmandi2013} have proposed the {\em Molecular Concentration Shift Keying (MCSK)} modulation scheme, where the main idea is to use the distinct molecule types for consecutive time slots at the transmitter, thus effectively suppressing the ISI. In \cite{Movahednasab2014} the authors take advantage of a limited memory in the transmitter to propose an adaptive transmission rate scheme that reduces ISI.   An on-off MOSK modulation scheme has been proposed in \cite{Kabir2014} which increases the transmission rate compared to MOSK but still suffers from the ISI. 
In \cite{Puda2014}, a modulation, called run-length hybrid aware, has been presented, which considers the runs (the same value occurs in several consecutive bits) for encoding: the run-value is encoded in molecule type and the run-length is encoded by CSK.
Though suffering from the ISI, this scheme is stated to improve the transmission rate in comparison with CSK and MOSK considering the ideal channel with no transmission errors. Authors in \cite{Burcu2015} propose a pre-equalization method to mitigate ISI where the signal at the receiver is considered as the difference between the number of the received molecules of each type while two types of molecules is used at the transmitter.  Further, several channel coding schemes are proposed to mitigate ISI in diffusion based MC \cite{Leeson2012, Ko2012, Shih2012} where they employ additional bits to intended input codewords in order to compensate errors from the ISI or noisy channel. 

\emph{Motivation for ISI-free modulations:} 
Given the generally negative effect of ISI, it is of interest to design modulation or channel coding schemes that would prevent or mitigate the ISI. To mitigate the ISI, the transmitter needs to make sure that molecules of the same type are released at time instances that are sufficiently far apart. Therefore, the transmissions will not have any superposition. This simplifies the receiver, who monitors the presence and intensity of various molecule types in the environment. Furthermore,  ISI-free modulation schemes are more robust when the receiver knows little about the channel state information (CSI). A high receiver molecule intensity can be either due to a high transmission amplitude or a very conductive channel. In lack of CSI and potential presence of ISI, observing a certain intensity of a molecule type, the receiver has to meet the challenge of deciding whether the received molecule density is due to superposition of various successive transmissions or due to temporary channel conductivness.

In this paper, we study ISI-free modulation schemes for time-slotted transmission model. The problem of finding the maximum achievable rate is converted into the design of a codebook, consisting of string of symbols, with each symobl indicating the type and concentration of a molecule transmission. The ISI-free condition implies that symbols of the same type should not be used close to each other. This problem can be expressed in terms of the constrainted coding problem. Constrained coding has found applications in storage and communication systems; its basic idea is to avoid transmission of input sequences that are more prone to error (e.g. see \cite[Chapter 1]{ConstrainedCoding}). Similarly, in our setup, we would like to avoid input sequences that cause ISI. 

Acheiving the constrained coding capacity requires blocklengths tending to infinity and complicated coding schemes. On the other hand, due to the limited resources available at nano-level, only \emph{simple} schemes are eligible for being implemented in practice. Further, we desire short blocklengths (limited delay) and modest memory resources. Our main contribution in this paper is to construct simple ISI-free modulation schemes for a given maximum transmission delay. It is shown (numerically) these simple schemes are near optimal, \emph{i.e.,} they become close to the constrained coding capacity.  

The rest of this paper is organized as follows. In Section \ref{secII}, ISI-free modulation is formally defined and modeled as a constraint graph. A summary of the notation used in this paper is given in Table \ref{Notation} at the beginning of this section. Section \ref{sec3} introduces the ISI-free modulation capacity under the limited delay (blocklength). In Section \ref{sec4} a class of modulation strategies based on a modified version of the constraint graph is proposed as a lower bound on the capacity. The numerical results and conclusions are provided in Section \ref{sec5} and \ref{sec6} respectively.

\section{ISI-free modulation problem for diffusion channels}\label{secII}
\begin{table}
\caption{Summary of the notation}
\begin{center}
\begin{tabular}{|c|c|}
\hline
Notation & Description\\ \hline
$k$ & Diffusion channel memory in terms of time slot\\ \hline
$N$ & The number of molecule types\\ \hline
$M_i$ & Symbol corresponding to molecule type $i$\\ \hline
$-$ & Symbol corresponding to no molecule transmission\\ \hline
$s_i$ & $i$-th symbol in a symbol sequence\\ \hline
$\textbf{s}$ & a symbol string or a state\\ \hline
$\ell$ & the length of a binary string\\ \hline
$m$ & length of a symbol string\\ \hline
$d$ & Delay (blocklength) constraint\\ \hline
$\mathcal{E}$ & Encoding function\\ \hline
$\mathcal{D}$ & Decoding function\\ \hline
$\mathcal{C} (d)$ & ISI-free capacity given blocklength $d$\\ \hline
$\mathcal{R}_\mathsf{L}(d)$ & Lower bound rate given blocklength $d$\\ \hline
$\mathcal{U}(\textbf{s})$ & Set of all possible uniquely decodable subsets\\ & corresponding to edges that exit from state $\textbf{s}$\\ \hline
$U(\textbf{s})$ & a member of set $\mathcal{U}(\textbf{s})$\\ \hline
$\mu_{U}(\mathbf{s})$ & a full prefix free binary code mapped to sequences in ${U}(\mathbf{s})$ \\ \hline  
\end{tabular}
\end{center}
\label{Notation}
\end{table}
Herein, we study ISI-free modulation schemes for the following setup: time is divided into slots of a fixed duration. The transmitter operates at the beginning of each time-slot: it either does not release any molecules, or chooses one molecule type and releases it into the environment. For simplicity of exposition, we assume that the information is coded \footnote{Please note that we use the words ``code" and ``modulation" and their derivatives interchangeably.}  into the molecule type only, and not in the number of released molecules (on-off keying modulation); extensions to amplitude modulation is also possible. Therefore, the transmission sequence can be expressed as a string of symbols, each of which is either a $-$ (no transmission), or $M_i$ when molecule of type $i$ is used. 

To avoid ISI at the receiver, there should be some time delay between two consecutive use of molecules of the same type. The amount of this delay depends on the communication medium. The diffusion channel is assumed to have $k$ time slot memory. Thus, we restrict the transmission sequence by requiring that there should be at least $k$ symbols between any two consecutive symbols $M_i$ for some parameter $k$. For instance, if $k=2$,   $M_1-M_2M_1$ and $M_1--M_1$ are valid sequences, but $M_1-M_1M_2$ is invalid. 

\begin{definition}A sequence of symbols $(s_1, s_2, s_3, \ldots, s_n)$ is called ISI-free if no the symbol of the form $M_r$ appears more than once in any consecutive substring of length $k+1$. The symbol, `$-$', is however, allowed to be repeated more than once. In other words, for any $1\leq i\leq n-k$, the substring $s_{i}, s_{i+1}, \ldots, s_{i+k}$, has the property that $s_j=s_r$ for $j\neq r\in \{i, i+1, \ldots, i+k\}$ only if $s_j=s_r=$`$-$'.
\end{definition}

We assume that the receiver has receptors that can detect the presence of any molecule type over time. Having avoided the ISI, the receiver can fully recover the transmission sequence of symbols. Then, the task of the transmitter is to encode information bits into a transmission sequence that obeys the restriction given above. 

\textbf{Example:}
An example of an ISI-free modulation is the MCSK modulation scheme,  proposed in \cite{Arjmandi2013}, for diffusion channels with one symbol memory ($k=1$).\footnote{ MCSK can be easily extended to the case of general channel with $k$ memory by using $k+1$ different molecule types.} MCSK uses the symbol set $M=\{-, M_1, M_2\}$ and prevents ISI by avoiding the transmission of the same molecule type in two consequent time slots. In binary MCSK, `0' input bit is mapped to `$-$' symbol; to transmit bit `1', in even time slots symbol $M_1$ is used while in the odd time slots, symbol $M_2$ is used. Note that quaternary and higher modulation orders can be developed by using more concentration levels of different molecule types. Fig.~\ref{Figure1} demonstrates the state diagram for the MCSK. In this figure, each edge is labeled by $x/y$, where $x$ and $y$ indicate the transmitted molecule and corresponding input bit, respectively. Observe that the MCSK scheme communicates at the rate of 1 bit per symbol.

\begin{figure}
\centering
\begin{tikzpicture}[->,>=stealth',shorten >=1pt,auto,node distance=6cm,
thick,main node/.style={circle,draw,font=\sffamily\small}]
\node[main node] (1) [align = center]{Odd \\ time slot};
\node[main node] (2) [right of=1,xshift=3cm,align=center] {Even \\ time slot};

\path[every node/.style={font=\sffamily\small}]
(1) edge  [bend left=40] node {$M_1/1$} (2)
(1) edge [bend left=20] node{\textemdash $/0$} (2)
(2) edge [bend left=20] node{$M_2/1$} (1)
(2) edge  [bend left=40] node{\textemdash $/0$} (1);

\end{tikzpicture}
\caption{State diagram of the MCSK modulation scheme}
\label{Figure1}
\end{figure}
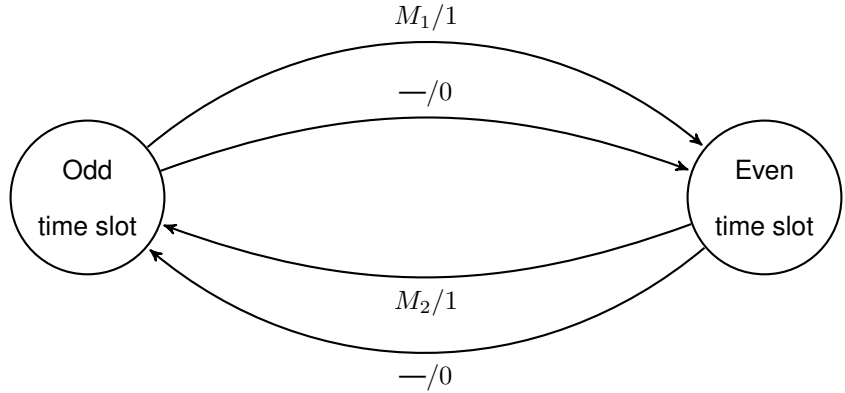
 \begin{figure}
\centering
\begin{tikzpicture}[->,>=stealth',shorten >=1pt,auto,node distance=6cm,
thick,main node/.style={circle,draw,font=\sffamily\small}]
\node[main node,text width=1cm] (1) [align = center]{$M_1$};
\node[main node,text width=1cm] (2) [above right of=1,align=center] {\textemdash};
\node[main node,text width=1cm] (3) [below right of=2,align=center] {$M_2$};

\path[every node/.style={font=\sffamily\small}]
(1) edge  [bend left=20] node{\textendash} (2)
	edge  [bend left=20] node {$M_2$}(3)
(3) edge  [bend left=20] node [right]{\textendash }(2)
      edge  [bend left=20] node{$M_1$} (1)
(2)  edge  [loop above=40] node{\textendash } (2)
      edge  [bend left=20] node [left]{$ M_1 $}(1)
      edge  [bend left=20] node{$ M_2 $} (3);
\end{tikzpicture}
\caption{State diagram of ISI-free sequences of the symbol set $M= \{- , M1, M2\}$ for channel with $k = 1$}
\label{Figure2}
\end{figure}
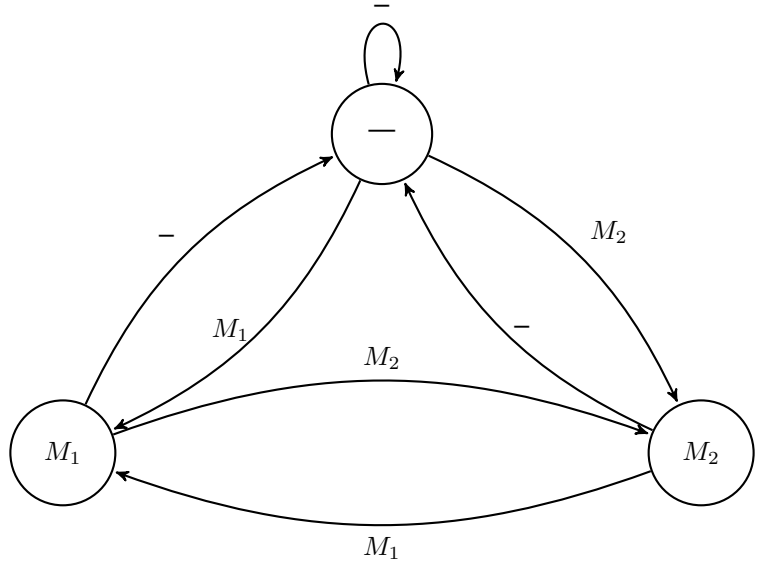

Observe that after transmission of the symbol `$-$', \emph{i.e.,} no molecule transmission, in fact all of the three symbols $M_1$ and $M_2$ and `$-$' can be transmitted without arising ISI. This degree of freedom is depicted in the state diagram of Fig.~\ref{Figure2}. If we move on the edges of this graph and write down the sequence of symbols written on the edges, we get an ISI-free transmission sequence. Conversely, any ISI-free sequence corresponds to a path on this graph (we allow an edge to occur arbitrarily many times in a path). 
 Therefore, the number of paths of length $m$ on this graph give us with the maximum number of messages that we can send by transmitting $m$ symbols. A transmission rate $R$ is equivalent with sending a binary sequence of length $mR$ using $m$ symbols, \emph{i.e.,} the number of messages is $2^{mR}$ for a code of length $m$. Therefore, we are interested in the logarithm of the number of paths in the graph, divided by $m$. 

When the channel memory $k$ and the number of molecule types are arbitrary, one can draw a state diagram similar to  Fig.~\ref{Figure2} as follows:
\begin{definition}[Constraint graph]\label{def:const-graph}
 Each vertex (state) is defined as the vector of the molecule types of the last $k$ transmitted symbols and each of the edges emanating from a vertex (state) represents an allowed symbol for transmission given the last $k$ transmitted symbols. In other words, the nodes of the graph are $k$-tuples $(s_1, s_2, \cdots, s_k)$ where $s_i=s_j$ for $i\neq j$ occurs only if $s_i=s_j=$`$-$'. There is a directed edge from vertex $(s_1, s_2, \cdots, s_k)$ to $(s_2, s_3, \cdots, s_k, s_{k+1})$, only if $(s_1, s_2, \cdots, s_k, s_{k+1})$ is an ISI-free sequence, \emph{i.e.}, either $s_{k+1}=$`$-$', or $s_{k+1}\neq s_1$. The edge from vertex $(s_1, s_2, \cdots, s_k)$ to $(s_2, s_3, \cdots, s_k, s_{k+1})$ is labled by the symbol $s_{k+1}$.
\end{definition}

The number of paths of  length $m$ on a directed graph (starting from a given node) is denoted by $N(m)$ and  is a simple combinatorical problem that can be found using recursive equations. The logarithm of the number of paths of length $m$ divided by $m$, as $m$ converges to infinity, can be readily found in terms of the adjacency matrix of the graph, as it is a special case of the classical problem of \emph{constrained coding} \cite[Theorem 3.4]{ConstrainedCoding}. 
\begin{theorem}(\cite[Theorem 3.4]{ConstrainedCoding})\label{thm:constr-coding} Let $\mathcal{G}$ be an irreducible constraint graph representing an irreducible constrained system. Then,
$$\underset{n\rightarrow \infty}{\lim}sup\frac{1}{m}\log N(m) = \log_2 \lambda(A_\mathcal{G}),$$
where $A_\mathcal{G}$ is the adjacency matrix of the graph $\mathcal{G}$ and $\lambda(A_\mathcal{G})$ is the largest of the absolute values of eigenvalues of $A_\mathcal{G}$.
\end{theorem}

The connection between constrained coding and ISI-free coding is known and studied in applications such as magnetic recording \emph{constrained coding} \cite[Chapter 1]{ConstrainedCoding}. The earliest results on constrained coding problem are due to  Shannon \cite{Shannon1948} who considered teletype and telegraphy systems as two examples of channels with restrictions on the input symbol sequences. A general result for transmission of a binary symmetric source over a finite state discrete noiseless channel is given by Shannon in \cite{Shannon1948}.

According to Theorem \ref{thm:constr-coding}, one can easily verify that the maximum achievable transmission rate for the graph of  Fig.~\ref{Figure2} is $C=1.2716$. The adjacency matrix for this graph would be as follows:
\[A_{\mathcal{G}}=
\begin{blockarray}{cccc}
- & M_1 & M_2 \\
\begin{block}{(ccc)c}
  1 & 1 & 1 & - \\
  1 & 0 & 1 & M_1  \\
  1 & 1 & 0 & M_2 \\
\end{block}
\end{blockarray}
\]
and for the given adjacency matrix $\lambda(A_\mathcal{G})=2.4142$, so the maximum achievable rate equals $\log_2(2.4142) = 1.2716$.  But achieving this rate requires long blocklength that implies long memory and transmission delay at the transmitter and receiver. Nonetheless, we now show that one can reach the transmission rate of $1.25$ with blocklength (delay) of one. We do this by slightly modifying the MCSK scheme (MCSK was transmitting at rate 1 bit per symbol). 

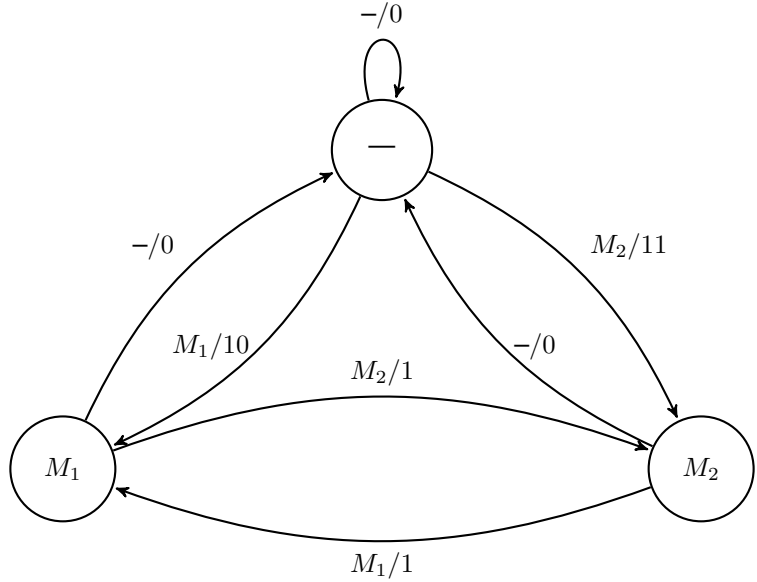
\begin{figure}
\centering
\begin{tikzpicture}[->,>=stealth',shorten >=1pt,auto,node distance=6cm,
thick,main node/.style={circle,draw,font=\sffamily\small}]
\node[main node,text width=1cm] (1) [align = center]{$M_1$};
\node[main node,text width=1cm] (2) [above right of=1,align=center] {\textemdash};
\node[main node,text width=1cm] (3) [below right of=2,align=center] {$M_2$};

\path[every node/.style={font=\sffamily\small}]
(1) edge  [bend left=20] node{\textendash $/0$} (2)
	edge  [bend left=20] node {$M_2/1$}(3)
(3) edge  [bend left=20] node [right]{\textendash $ /0$}(2)
      edge  [bend left=20] node{$M_1/1$} (1)
(2)  edge  [loop above=40] node{\textendash $ /0$} (2)
      edge  [bend left=20] node [left]{$ M_1/10 $}(1)
      edge  [bend left=20] node{$ M_2/11 $} (3);
\end{tikzpicture}
\caption{A modulation strategy for the diffusion channel and
the symbol set $M= \{- , M_1, M_2\}$ for channel with $k = 1$.}
\label{Figure3}
\end{figure}

Consider the modulation scheme given in Fig.~\ref{Figure3} with the same $x/y$ convention as before. Here, for instance if we are in state $-$, we send symbol $-$ if the next message bit we want to transmit is zero; we send $M_1$ if the next two bits are $10$ and  we send $M_2$ if the next two bits are $11$.  In this modulation scheme, the average transmission rate in states $-, M_1, M_2$ are $1.5, 1, 1$ bits per symbol respectively. Also, the probability of being in states $-, M_1, M_2$  can be computed using corresponding linear equations over the states, which will be equal to ${0.5, 0.25, 0.25}$, respectively. Therefore, this scheme achieves the rate of $1.25$.

The implication of the above construction is the existence of near optimal modulation strategies that are efficient in terms of blocklength and decoding delay, as the rate $1.25$ is very close to the constrained coding limit of $1.2716...$. The main result of this paper is constructing simple memory-limited ISI-free modulation schemes for a given maximum transmission delay that achieve rate close to the constrained coding limit.

\section{ISI-free modulations with limited delay and decoder buffer size}\label{sec3}

In the previous section, we observed that the ISI-free capacity, when there is no restriction over  delay, is equivalent to the constrained coding capacity. In practical modulation schemes for MC, due to limitations on memory resources and the fact that  large decoding delays may not be allowed in some applications, we are interested in assigning the input bits to ISI-free sequences of symbols with a limited length, \emph{i.e.}, using codes with a short blocklength. 

\begin{definition}[‎Delay (blocklength) constraint] Delay constraint $d$ means that the decoding of the $i$-th bit of the input binary string, $b_i$, can be delayed up to $d$ channel uses. More precisely, in using a variable length code for transmission of the input data bits, no codeword of length more than $d$ should be used. This implies that upon the transmission of the bit  $b_i$, it will be decoded within $d$ time-slots.
\end{definition} 

An ISI-free code of length $n$, with a given blocklength (delay) constraints consists of 
\begin{itemize}
\item An encoder  function, $\mathcal E$, mapping the input bit string $\{b_1, \cdots , b_n\}$ to an ISI-free sequence of the symbols $(s_1,\cdots, s_m)$. The number of output symbols $m$ may depend on  the input bit string. The rate of the code is $R=n/\mathbb{E}[m]$ (bits per symbol).

The maximum blocklength $d$ constraint implies the following structure on the encoding function $\mathcal E=(\mathcal E_1, \mathcal E_2, \cdots)$: assuming that bits $(b_{1}, b_{2}, \cdots, b_{l_{i}})$ are already encoded into symbol string $(s_1, s_{2}, \cdots, s_{\tau_i})$, the encoder $\mathcal E_i$ takes in new bits $b_{l_{i}+1},\ldots, b_{l_{i+1}}$
and maps it to output symbols $( s_{\tau_i+1}, s_{\tau_i+2}, \cdots, s_{\tau_{i+1}})$ of length less than $d$, \emph{i.e.,} $\tau_{i+1}-\tau_i\leq d$. Observe that $l_i$ is the number of input bits that are encoded in the first $i$-th steps and $\tau_i$ is the number symbols that are transmitted during this step.

\item A decoder function $\mathcal{D}_i$ at step $i$ that reverses the encoding operation by taking in symbols $(s_1, s_2, \cdots, s_{\tau_{i+1}})$ and producing bits $b_{l_{i-1}+1},\ldots, b_{l_i}$ where $l_i$ is the number of input bits that are decoded in the $i$-th step. 
\end{itemize}

Given $d$, the ISI-free capacity of the system is defined as the maximum rate that is asymptotically achievable:
\begin{align}
 \mathcal{C}(d)= \max_{n, \mathcal{E}_i, \mathcal{D}_i}~\frac{n}{\mathbb{E}[m]}.
 \label{rate}
\end{align}

It is clear that $\mathcal{C}(d)\leq \mathcal{C}(d')$ if $d\leq d'$.  Further, the maximum achievable rate of the system when we do not have any  delay constraint, $\mathcal{C}(\infty)$, is equal to the constrained coding capacity. Therefore, constrained coding capacity is always an upper bound for $\mathcal{C}(d)$. In this paper, we propose a suboptimal approach to determine the encoder and decoder function for ISI-free modulation problem under  delay constraints. The associated rate gives a lower bound on $\mathcal{C}(d)$, which we denote by  $\mathcal{R}_\mathsf{L}(d)$. Thus,
$$\mathcal{R}_\mathsf{L}(d)\leq \mathcal{C}(d) \leq \mathcal{C}(\infty).$$

Simulation results show that $\mathcal{R}_\mathsf{L}(d)$ is close to $\mathcal{C}(\infty).$ for small values of  $d$. As a result, $\mathcal{C}(d)$ is not far from $\mathcal{C}(\infty).$ and the proposed scheme is near optimal. To propose the lower bound, we restrict to a class of modulation schemes constructed based on state diagram of ISI-free symbol sequences.

\begin{remark}
Although our aim is to study diffusion molecular channels, but the delay-limited codes that we construct  in the next section can be used in other similar constrained coding applications.
\end{remark}

\section{Lower bound on $\mathcal{C}(d)$}\label{sec4}
Consider a diffusion channel with $k$  symbol memory, where the symbols are selected from the set  $M=\{ -, M_1,...,M_N\}$,  in which $N$ is the number of molecule types available to the transmitter.  In an ISI-free modulation, no molecule type appears repeatedly, in each sequence of output symbols of  length $k+1$.\footnote{The number of  molecule types $N$ may be less than $k$, since we are allowed to use $-$ repeatedly in constructing the ISI-free sequences.} As discussed earlier, the ISI-free sequences can be represented as a path on the constraint graph of Definition \ref{def:const-graph}. One example of such a graph is given in Figure \ref{Figure2}. To construct a good code, we used the graph given in Figure \ref{Figure3}, reaching the performance of $1.25$ which is close to the optimal $1.2716$. 

The main idea of this section is to provide a systematic algorithm for constructing codes similar to the one given in  Figure \ref{Figure3} from the constraint graph of  Figure \ref{Figure2}. To do this, we first define a modified  \emph{constraint graph of maximum depth $d$}  to represent a practical ISI-free modulation family under the memory and delay constraints. We then discuss assigning input bits to edges of this modified constraint graph. 

In the modified constraint graph, the vertices are not changed  but a permissible sequence of maximum length $d$ (instead of a single symbol) is assigned to each edge. 
\begin{definition}[Constraint graph of depth $d$]\label{def:const-graph-Z}
Let $d$ be a natural number. The nodes of the constraint graph of depth $d$ are $k$-tuples $(s_1, s_2, \cdots, s_k)$ of permissible sequences, \emph{i.e.,} $s_i=s_j$ for $i\neq j$ only if $s_i=s_j=$`$-$'. Given any natural number $t\leq d$ and any ISI-free string $(s_1, s_2, \cdots, s_k, \cdots, s_{k+t})$, we draw a directed edge from vertex $(s_1, s_2, \cdots, s_k)$ to $(s_{t+1}, s_{t+2}, \cdots,  s_{t+k})$. This edge is labeled by the string $(s_{k+1}, s_{k+2}, \cdots, s_{t+k})$.
\end{definition}

 Different ISI-free modulations are obtained by assigning sequences of  binary bits to a set of edges emanating from each state. The  modulator produces the ISI-free sequence by moving over the introduced graph based on the input bit sequence and transmitting the corresponding symbols. The demodulator reconstructs the bit sequence by  moving on the same graph based on the received symbol sequences. In this modulation strategy, the demodulator needs to buffer at most the last $d+k$ received symbols, imposing a limited symbol buffer of length $d+k$ at the receiver.
 
More accurately, given a state $\mathbf{s}=(s_1, s_2, \cdots, s_k)$, the emanating edges from the state are labeled with a string of length  $t\leq d$, $s_{k+1}, s_{k+2}, \ldots, s_{k+t}$ such that $(s_1, s_2, \cdots, s_{k+t}$ is ISI-free. We would like to select a subset of these edges and assign  binary bits to them. This assignment specifies the edge we would choose for moving, based on the input bit string. The only restriction is that the set of sequences $s_{k+1}, s_{k+2}, \ldots, s_{k+t}$ of the edges that we choose must be prefix-free (uniquely decodable). This would enable the possibility of recovering the input bit string from the set of output strings. 
\begin{example}\label{example1}
For the symbol set $M=\{-,M_1,M_2\}$, $d=2$ and $k=1$, the constraint graph of depth $d$ has three states $-, M_1$ and $M_2$. Given the state $\mathbf{s}=M_1$, the allowed edges emanating from this state are labeled by $\{-, M_2, --,-M_1,-M_2, M_2-, M_2M_1\}$. From these seven edges that leave state $M_1$, there are two edges (labeled by $M_2$ and $-M_2$) that go from state $M_1$ to state $M_2$; there are two edges (labeled by $-M_1$ and $M_2M_1$) that go from state $M_1$ to itself; and there are three edges (labeled by $-$, $--$, and $M_2-$) that go from state $M_1$ to $-$. Among these seven edges, the five edges $\{--,-M_1,-M_2, M_2M_1,M_2-\}$ form a possible uniquely decodable sequence set (no members of this set is a prefix of the another).  
\end{example}

Given a state $\mathbf{s}$, let $\mathcal{U}(\mathbf{s})$ denote the set of all possible uniquely decodable subsets corresponding to edges that exit from state $\mathbf{s}$. To come up with an explicit modulation, for each state $\mathbf{s}$, the encoder parses the input data bits by a full prefix-free code $\mu_{U}(\mathbf{s})$ and maps each codeword to a sequence  from the $U(\mathbf{s})\in \mathcal{U}(\mathbf{s})$.

\begin{example}Consider Example \ref{example1}, the state  $\mathbf{s}=M_1$ and $U(\mathbf{s})=\{--,-M_1,-M_2, M_2M_1,M_2-\}$. Given that $U(\mathbf{s})$ has five sequences, to define $\mu_{U}(\mathbf{s})$ for this state, we may pick any Huffman code on five symbols, \emph{e.g.} $\{0, 10, 110, 1110, 1111\}$, and assign it to the sequences of $U(\mathbf{s})$ respectively. Thus, if we are at state 
$\mathbf{s}=M_1$ and the next input bit is $0$, we move along edge $--$; if the next two input bits are $10$, we move along $-M_1$, etc.
\end{example}
Assuming that $U(\mathbf{s})=\{\mathbf{s}_1, \mathbf{s}_2, \cdots, \mathbf{s}_r\}$ and we assign a binary Huffman code with string of length $\ell_i$ to sequence $\mathbf{s}_i$, we must have  $\sum_{i=1}^{r}{2^{-\ell_i}}=1$. Furthermore, the probability that we take the edge along the sequence  $\mathbf{s}_i$ will be $2^{-\ell_i}$, as the input is assumed to be a uniform and independent string of bits. If the length of  $\mathbf{s}_i$ is $m_i$ symbols, the average transmission rate (bit per symbol) given the state $\mathbf{s}$ will be equal to
\begin{align}
R_{\mathbf{s}}=\frac{\mathbb{E}_{\mathbf{s}}(\ell)}{\mathbb{E}_{\mathbf{s}}(m)}=\frac{\sum_{i=1}^{r}{2^{-\ell_i}\ell_i}}{\sum_{i=1}^{r}{2^{-\ell_i}m_i}}\label{eqn2|}
\end{align}
in which $\mathbb{E}_{\mathbf{s}}(\ell)$ and $\mathbb{E}_{\mathbf{s}}(m)$ denote the average number of transmitted bits and symbols given the state $\mathbf{s}$.  The overall achievable rate by this scheme for large input bit stream can be computed by 
\begin{align}
R=\frac{\sum_{\mathbf{s}}p(\mathbf{s})\mathbb{E}_{\mathbf{s}}(\ell)}{\sum_{\mathbf{s}}p(\mathbf{s})\mathbb{E}_{\mathbf{s}}(m)}
\end{align}
where $p(\mathbf{s})$ denotes the stationary probability of the state $\mathbf{s}$ and these probabilities are computed using recursive linear equations of states probabilities.

\textbf{Lower bound on $\mathcal{C}(d)$:} Any coding scheme that satisfies the delay constraint $d$ would yield a lower bound to $\mathcal{C}(d)$. Consider a code constructed based on the constraint graph of depth $d$; this code will satisfy the delay constraint of $d$ because all the codewords on edges of the graph are of length at most $d$ symbols.  The optimal strategy is determined by the uniquely decodable sequence set and full prefix code given each state that achieves the maximum average rate (for large input bit streams) under limited blocklength constraint,\emph{i.e.},
\begin{align}\label{LBprob}
\mathcal{R}_\mathsf{L}(d)=\max_{U(\mathbf{s}), \mu_{U}(\mathbf{s}) \text { for all } \mathbf{s}}~\frac{\sum_{\mathbf{s}}p(\mathbf{s})\mathbb{E}_{\mathbf{s}}(\ell)}{\sum_{\mathbf{s}}p(\mathbf{s})\mathbb{E}_{\mathbf{s}}(m)}.
 \end{align}
The above optimization problem involves maximizing a ratio. The lower bound $\mathcal{R}_\mathsf{L}(d)$ can be expressed in an alternative \emph{linear form}, which makes it amenable to dynamic programming analysis: let 
\begin{align}G(R)&=\max_{U(\mathbf{s}), \mu_{U}(\mathbf{s})}\nonumber
\sum_{\mathbf{s}}p(\mathbf{s})\mathbb{E}_{\mathbf{s}}(\ell)-R\cdot\sum_{\mathbf{s}}p(\mathbf{s})\mathbb{E}_{\mathbf{s}}(m)
\\&=\max_{U(\mathbf{s}), \mu_{U}(\mathbf{s})}\sum_{\mathbf{s}}p(\mathbf{s})\big(\mathbb{E}_{\mathbf{s}}(\ell)-R\cdot \mathbb{E}_{\mathbf{s}}(m)\big).\label{G-formII}\end{align}
Then $G(R)$ is a monotonically strictly decreasing function, with $G(0)>0$. Furthermore, $G(\mathcal{R}_\mathsf{L}(d))=0$. Therefore, the value of $\mathcal{R}_\mathsf{L}(d)$ can be found as the unique positive root of $G(R)$. An efficient algorithm for computing $G(R)$ would lead to an efficient algorithm for approximating $\mathcal{R}_\mathsf{L}(d)$, by employing the bisection method for computing the root of the strictly decreasing function $G(R)$.

\textbf{Dynamic programming (DP) algorithm:} The expression for $G(R)$ in equation \eqref{G-formII} is in a form that can be solved using DP. The state space for the DP are the states $\mathbf{s}$ (the nodes of the graph). Let us interpret $\big(U(\mathbf{s}), \mu_{U}(\mathbf{s})\big)$ as the \emph{action} that we take at state $\mathbf{s}$. The expression for $\mathbb{E}_{\mathbf{s}}(\ell)-R\cdot \mathbb{E}_{\mathbf{s}}(m)$ depends only on the action $\big(U(\mathbf{s}), \mu_{U}(\mathbf{s})\big)$ that we take in  state $\mathbf{s}$; it does not depend on $p(\mathbf{s})$. Therefore, $\mathbb{E}_{\mathbf{s}}(\ell)-R\cdot \mathbb{E}_{\mathbf{s}}(m)$ can be viewed as the one-step profit, when we take the action $\big(U(\mathbf{s}), \mu_{U}(\mathbf{s})\big)$ at the state $\mathbf{s}$. And $G(R)$ will be the overall average profit of the DP. The state transition is defined in the natural way: upon taking the action  $\big(U(\mathbf{s}), \mu_{U}(\mathbf{s})\big)$ at the state $\mathbf{s}$, an outgoing edge corresponding to members of $U(\mathbf{s})$ is chosen according to the probability distribution  given by the length of the binary strings in $\mu_{U}(\mathbf{s})$; we move along this edge to get to a new state. 

The average profit $G(R)$ can be found iteratively using the standard dynamic programming algorithm. We now briefly explain an idea to use the structure of our particular DP to simplify its implementation: suppose we in an iteration of the DP algorithm. Fix some state  $\mathbf{s}$. In each iteration of the DP algorithm, we need to maximize (over all actions) the average one-step profit plus the profit of the state will end up with. We can do the search over $U(\mathbf{s})$ exhaustively. Then the  maximizing $\mu_{U}(\mathbf{s})$ can be found using the Geometric Huffman Coding approach \cite{Bochere2011}. To see this, observe that the average one-step profit plus the profit of the state we reach can be written as follows (using the expressions given in equation \eqref{eqn2|}):
\begin{align}
\big(\sum_{i=1}^{r}{2^{-\ell_i}\ell_i-R}\sum_{i=1}^{r}{2^{-\ell_i}m_i}\big) + \sum_{i=1}^{r}{2^{-\ell_i}c(\mathbf{s}_i)\label{on-s}}
\end{align}
where $c(\mathbf{s}_i)$ is the current state cost of the state $\mathbf{s}_i$ that we reach from state $\mathbf{s}$ with probability $2^{-\ell_i}$. The goal is to maximize \eqref{on-s} over all natural numbers $(\ell_1, \ell_2, \ldots, \ell_r)$ satisfying $\sum_{i}2^{-\ell_i}=1$. But equation \eqref{on-s} can be expressed as the KL divergence $-D(p\|q)$ for 
$p(i)=2^{-\ell_i}$ and $q(i)=2^{Rm_i-c(\mathbf{s}_i)}$. Observe that $q(\cdot)$ is not necessarily a probability distribution. Maximizing $-D(p\|q))$ or equivalently minimizing $D(p\|q)$ over diadic pmf $p$ is the problem considered in \cite{Bochere2011} and can be solved using the Geometric Huffman Coding approach.

\section{Numerical results for the diffusion channel}\label{sec5}

In this section numerical results are presented, indicating the near-optimality of the proposed modulation scheme. The numerical results illustrates how the system parameters affect the achievable bit rates. The parameters investigated here are the number of molecule types exploited in the system $N$, the channel memory $k$, and the maximum tolerable decoding delay $d$.

\begin{figure*}
\begin{center}
\includegraphics[scale=0.55,angle=0]{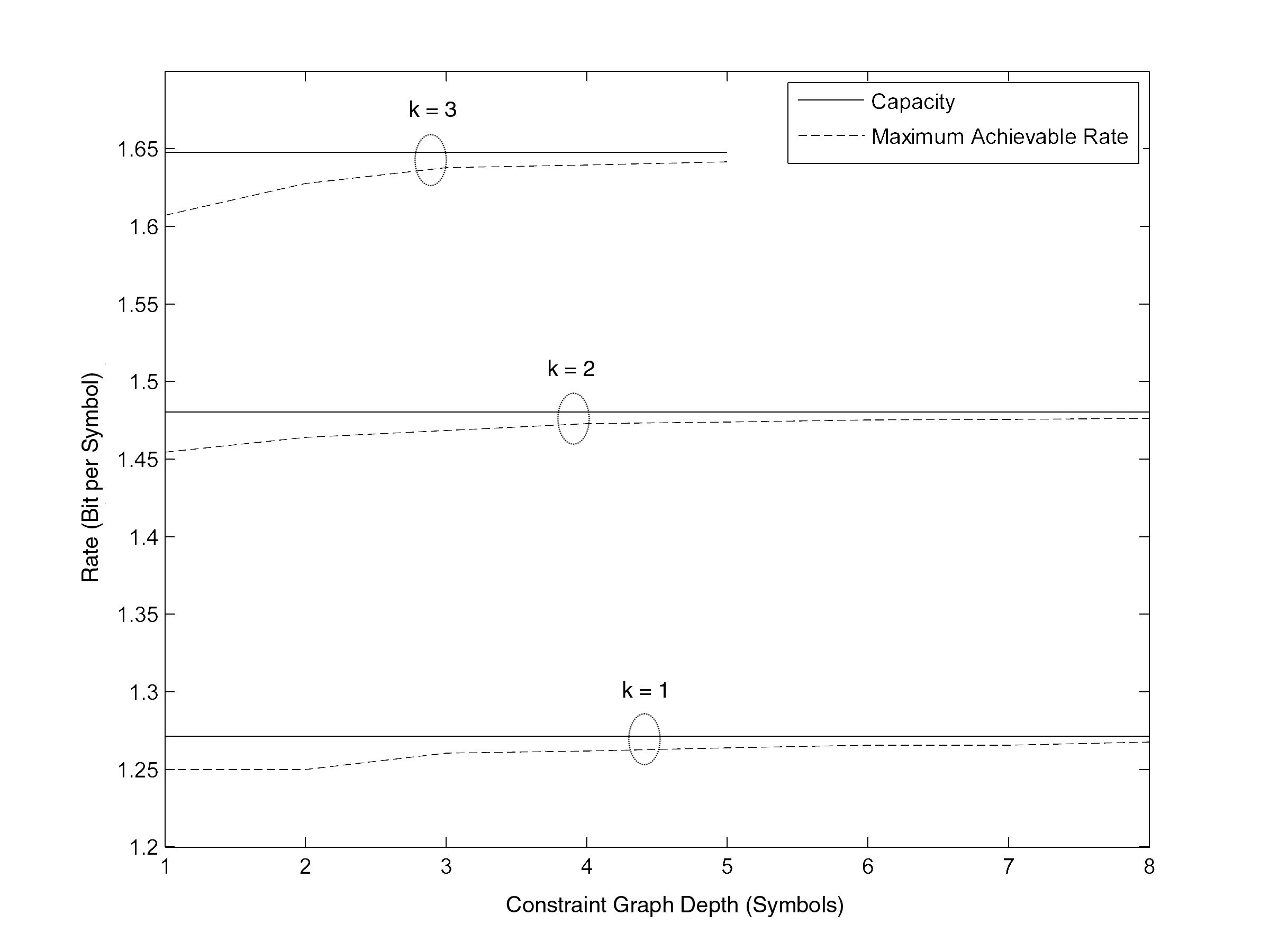}
\end{center}
\caption{ The effect of the blocklength (constraint graph depth) on the maximum achievable rate.}
\label{RateVsZ}
\end{figure*}

In Fig.~\ref{RateVsZ}, the achievable rates are plotted versus the constraint graph depth (\textit{i.e.} delay $d$). The figure compares the maximum achievable rate for a delay $d$. The curve is obtained by the dynamic programing (DP) algorithm. This figure is plotted for diffusion channels with up to three symbols memory, \textit{i.e.} $k=3$. For each channel, the number of molecule types are set to $k+1$, i.e, $N=k+1$. As we see in the figure, small values of $d$ are enough to approach the capacity.

\begin{figure*}
\begin{center}
\includegraphics[scale=0.55,angle=0]{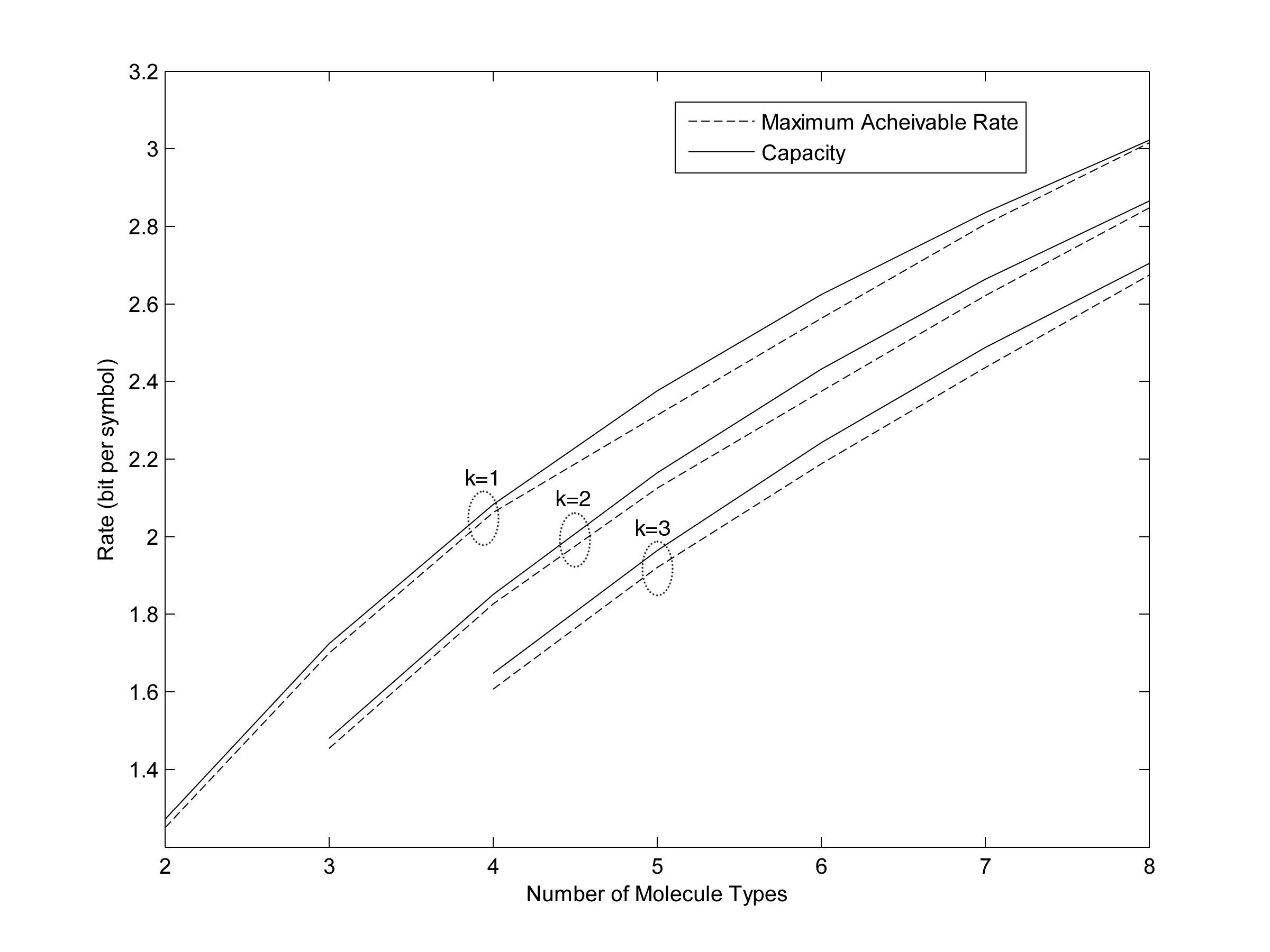}
\end{center}
\caption{ Maximum achievable rate vs number of molecule types for $d=1$ and $k=1,2,3$.}
\label{RateVsM}
\end{figure*}

In Fig.~\ref{RateVsM} the effect of number of molecule types on achievable rates is studied. Similar to Fig.~\ref{RateVsZ}, the achievable rates are obtained by the DP algorithm and are plotted for channels with up to three symbols memory $k=3$. For all the plots in this figure the depth of the constraint graph is set to a single symbol, $d=1$. It is observed that as the number of molecule types increases, the achievable rate increases. It should be noticed that in  Fig.~\ref{RateVsZ}, unlike the Fig.~\ref{RateVsM}, channels with higher memory have larger rates. The reason is that, the number of molecule types are not the same for all the plots as the number of molecule types $N$ increases \mbox{with $k$}.

\begin{figure*}
\begin{center}
\includegraphics[scale=0.55,angle=0]{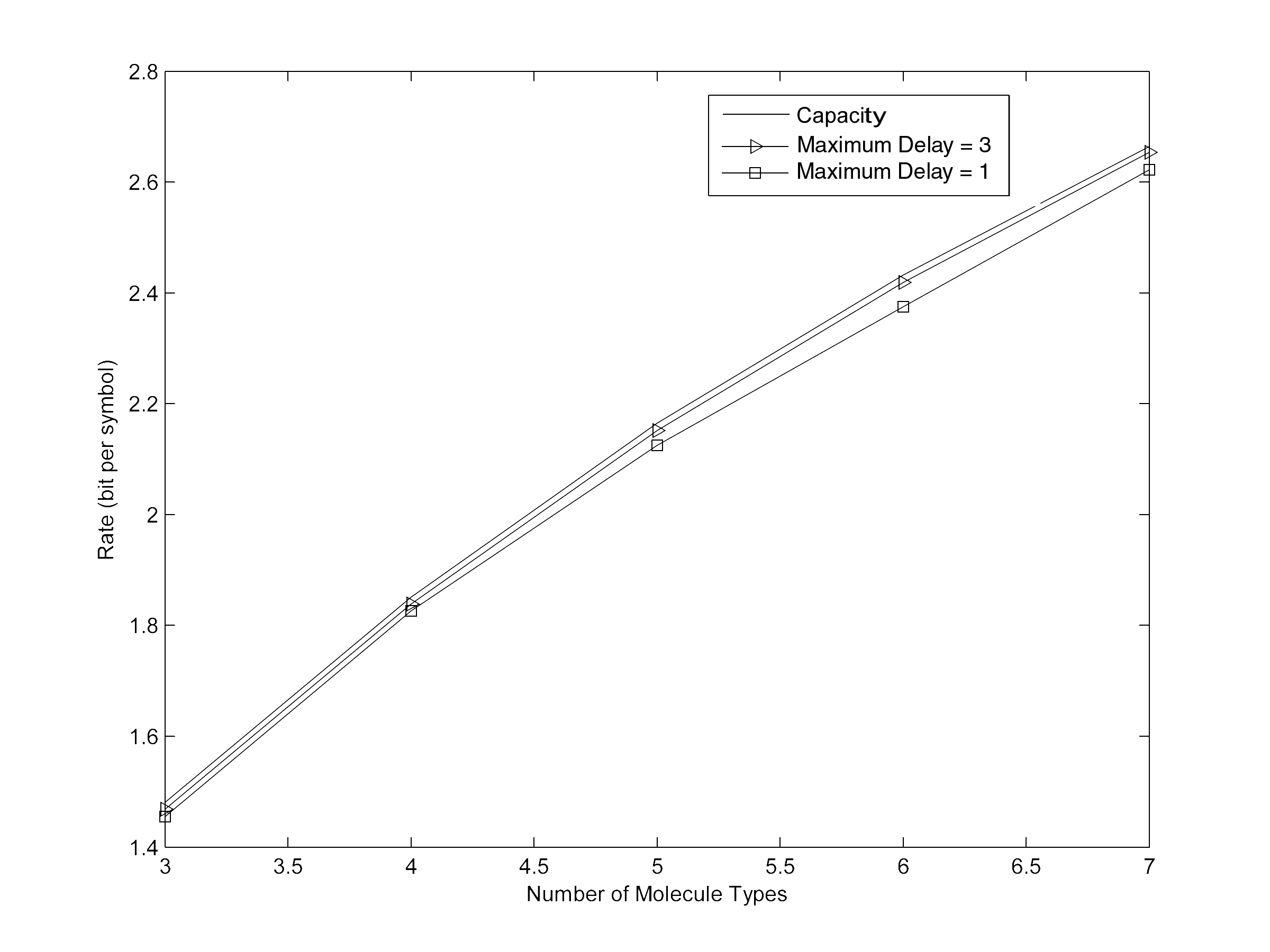}
\end{center}
\caption{ Maximum achievable rate vs number of molecule types for $k=1$ and $d=1, 3$.}
\label{RateVsM_K_fixed}
\end{figure*}

In Fig.~\ref{RateVsM_K_fixed} the achievable rates are  plotted versus the number of molecule types but for different values of delays $d$. As we see in this figure by increasing the depth from one to three, we can achieve the capacity for any number of molecule types but the gap occurs starting around  $N=5$.

Remember that MCSK was an ISI-free modulation for $k=1$ and $N=2$ from the symbol set $\{-,M_1,M_2\}$. Table \ref{ResT1} shows the achievable rate of different ISI free modulation schemes for $k=1$, $N=2$ and different values of $d$. As $d$ increases the optimal modulation found over constrained graph gives higher rate very close to the capacity. Observe that the rate obtained for $d=2$ is the same as that for $d=1$. Table \ref{ResT} shows the optimal code that one gets for the case of $d=2$, achieving the rate 1.25.

\begin{table}
\caption{Achievable rate for different ISI fee modulation schemes and different constraint graph depths.  $k=1$, $N=2$ and different $d$ values.}
\begin{center}
\begin{tabular}{|c|c|}
\hline
Modulation scheme & Achieved Rate\\ \hline
MCSK & 1\\ \hline
Optimal over constraint graph with $d=1$ & \\
(Modified MCSK) & 1.25\\ \hline
Optimal over constraint graph with $d=2$ &\\
(Table \ref{ResT}) & 1.25\\ \hline
Optimal over constraint graph with $d=3$ & 1.2604\\ \hline
Optimal over constraint graph with $d=4$ & 1.2617\\ \hline
Optimal over constraint graph with $d=5$ & 1.2640\\ \hline
Capacity for $d=\infty$ & 1.2716 \\ \hline
\end{tabular}
\end{center}
\label{ResT1}
\end{table}
\begin{table}
\caption{Optimized modulation for transmitting a uniform binary string over a diffusion channel with one symbol memory, according to a constraint graph with depth 2 and the symbol set ${-, M_1, M_2}$.}
\begin{center}
\begin{tabular}{|c|c|c|c|c|c|}
\hline
\multicolumn{2}{|c|}{State: $-$} & \multicolumn{2}{|c|}{State: $M_1$} & \multicolumn{2}{|c|}{State: $M_2$}\\ \hline
Symbol Seq. & Bit String & Symbol Seq. & Bit String & Symbol Seq. & Bit String \\ \hline
$-,-$          & 0,0,0  & $-,-$           & 0,0    & $-,-$          & 0,0\\
$M_1,-$     & 0,0,1  & $M_2,-$      & 0,1    & $M_1,-$      & 0,1\\
$M_2,-$     & 0,1     & $-,M_1$      & 1,0,0 & $-,M_1$      & 1,0,0\\
$-,M_1$     & 1,0,0  & $-,M_2$      & 1,0,1 & $-,M_2$      & 1,0,1\\
$-,M_2$     & 1,0,1  & $M_2,M_1$ & 1,1    & $M_1,M_2$ & 1,1\\
$M_1,M_2$ & 1,1,0 &                   &         &                   & \\
$M_2,M_1$ & 1,1,1 &                   &         &                   & \\
\hline

\end{tabular}
\end{center}
\label{ResT}

\end{table}

\section{Conclusion}\label{sec6}
We considered the problem of efficient rate ISI-free modulation in molecular communication under  limited  delay constraints. The ISI-free condition was expressed by a constrained graph where maximum achievable rate of modulations constructed over this graph, i.e, ISI-free capacity, is well known in constrained coding literature. Motivated by limited resources in nanomachines, we then defined the ISI-free capacity under  delay constraints, $\mathcal{C}(d)$. Afterwards, a suboptimal approach to determine the encoder and decoder functions (modulation scheme) that gives a lower bound on $\mathcal{C}(d)$ was proposed. The results show that these simple schemes are near optimal, i.e, become close to the constrained coding capacity.       

\begin{thebibliography}{21}


\bibitem{Akyl2011}
I. F. Akyildiz, J. M. Jornet, and M. Pierobon, ``Nanonetworks: A new frontier in communications,"  ‎\textit{Communications of the ACM}‎, vol. 54, no. 11, pp. 84-89, November 2011.

\bibitem{Nakano2012}
T. Nakano, M.J. Moore, F. Wei, A.V. Vasilakos, J. Shuai, ``Molecular communication and networking: Opportunities and challenges," ‎\textit{IEEE Transactions on NanoBioscience}‎, vol. 11, no. 2, pp. 135-148, 2012.
%



\bibitem{B2}
Moore, Michael J and Suda, Tatsuya and Oiwa, Kazuhoro, ``Molecular communication: modeling noise effects on information rate",  \textit{IEEE Transactions on NanoBioscience}, Vol. 8, No. 2, 2009.
%
\bibitem{Atakan2012}
B. Atakan, S. Galmes, and OB. Akan, ``Nanoscale communication with molecular arrays in nanonetworks." ‎\textit{IEEE Transactions on NanoBioscience}‎,  vol. 11, no. 2, pp. 149-160, 2012.

\bibitem{Pierobon2014}
M. Pierobon and I. F. Akyildiz, ``A statistical-physical model of interference in diffusion-based molecular nanonetworks." ‎\textit{IEEE Transaction on Communications}‎, vol. 62, no. 6, 2014.
\bibitem{Aminian2014}
G. Aminian, H. Arjmandi, A. Gohari, M. Nasiri Kenari U. Mitra, ``Capacity of Diffusion based Molecular Communication Networks in the LTI-Poisson Model," \textit{IEEE International  Conference on Communications}, June 2015.



\bibitem{Kuran2011}
M. S. Kuran, H. B. Yilmaz, T. Tugcu, I. F. Akyildiz, ``Modulation techniques for communication via diffusion in nanonetworks," \emph{IEEE International  Conference on Communications}, June 2011.

\bibitem{fekri}
A. Einolghozati, M. Sardari, F. Fekri, ``Design and Analysis of Wireless Communication Systems Using Diffusion-Based Molecular Communication Among Bacteria,", \textit{IEEE Transactions on  Wireless Communications}, vol.12, no.12, pp.6096,6105, December 2013.

\bibitem{Arjmandi2013}
H. Arjmandi, A. Gohari, M. Nasiri-Kenari and Farshid Bateni, ``Diffusion based nanonetworking: A new modulation
technique and performance analysis," ‎\textit{IEEE Communications Letters}‎, vol. no. 4, pp. 645 - 648, 2013.


\bibitem{Movahednasab2014}
M. Movahednasab, M. Soleimanifar, A. Gohari, M. Nasiri Kenari U. Mitra, ``Adaptive Molecule Transmission Rate for Diffusion Based Molecular Communication," \textit{IEEE International  Conference on Communications}, June 2015.

\bibitem{Kabir2014}
Md. Kabir, S. M. Islam, and K.S. Kwak. ``D-MoSK Modulation in Molecular Communications." \textit{IEEE Transactions on NanoBioscience}. 

\bibitem{Puda2014}
Pudasaini, Subodh, Seokjoo Shin, and Kyung Sup Kwak. ``Run-length aware hybrid modulation scheme for diffusion-based molecular communication," \textit{14th International Symposium on Communications and Information Technologies (ISCIT)}, 2014.

\bibitem{Burcu2015}
Tepekule, Burcu, et al. ``Novel Pre-Equalization Method for Molecular Communication via Diffusion in Nanonetworks," \textit{IEEE Communications Letters,} June 2015. DOI:10.1109/LCOMM.2015.2441726

\bibitem{Leeson2012}
M. S. Leeson, D. H. Matthew, ``Forward error correction for molecular communications," ‎\textit{Nano Communication Networks},‎ vol. 3, no. 3 , pp. 161-167, 2012.
\bibitem{Ko2012}
P. Y. Ko, et al. ``A new paradigm for channel coding in diffusion-based molecular communications: Molecular coding distance function," ‎\textit{Global Communications Conference (GLOBECOM)}‎, 2012.
\bibitem{Shih2012}
Shih, Po-Jen, Chia-han Lee, and Ping-Cheng Yeh. ``Channel codes for mitigating intersymbol interference in diffusion-based molecular communications," ‎\textit{Global Communications Conference (GLOBECOM)}‎, 2012.

\bibitem{ConstrainedCoding}
B. H. Marcus, R. M. Roth, P. H. Siegel, ``An Introduction to Coding for Constrained Systems," available at 
\url{ https://www.math.ubc.ca/~marcus/Handbook}

\bibitem{Shannon1948}
C. E. Shannon, ``A Mathematical Theory of Communication," \textit{The Bell System Technical Journal}, 1948.
\bibitem{Bochere2011}
G. Bocherer and R. Mathar, ``Matching dyadic distributions to channels," \textit{IEEE Data Compression Conference (DCC)}, 2011.


\end{thebibliography}

\end{document}